\documentclass[aps,preprint,groupedaddress]{revtex4}

\usepackage{graphics,epsfig}

\def\bit{\begin{itemize}}
\def\eit{\end{itemize}}
\def\ben{\begin{enumerate}}
\def\een{\end{enumerate}}

\def\beq{\begin{equation}}
\def\eeq{\end{equation}}
\def\bea{\begin{eqnarray}}
\def\eea{\end{eqnarray}}

\newenvironment{Eqnarray}%
     {\arraycolsep 0.14em\begin{eqnarray}}{\end{eqnarray}}

\def\beqa{\begin{Eqnarray}}
\def\eeqa{\end{Eqnarray}}
\def\beqno{\begin{eqalignno}}
\def\eeqno{\end{eqalignno}}
\def\ifmath#1{\relax\ifmmode #1\else $#1$\fi}
\def\lsim{\mathrel{\raise.3ex\hbox{$<$\kern-.75em\lower1ex\hbox{$\sim$}}}}
\def\gsim{\mathrel{\raise.3ex\hbox{$>$\kern-.75em\lower1ex\hbox{$\sim$}}}}

\def\anti{\overline}
\def\rtwo{\sqrt 2}

\def\mud{M_U}
\def\mdd{M_D}

\def\cbma{c_{\beta-\alpha}}

\def\sbma{s_{\beta-\alpha}}

\def\beqno{\begin{eqalignno}}
\def\eeqno{\end{eqalignno}}
\def\beq{\begin{equation}}
\def\eeq{\end{equation}}

\def\ifmath#1{\relax\ifmmode #1\else $#1$\fi}

\def\call{{\cal L}}
\def\tb{t_{\beta}}
\def\tbi{t_\beta^{-1}}

\def\sb  {s_{\beta}}
\def\cb  {c_{\beta}}

\def\sa  {s_{\alpha}}
\def\ca  {c_{\alpha}}

\def\hl{h^0}
\def\ha{A^0}
\def\hh{H^0}
\def\hpm{{H^\pm}}

\def\hp{{H^+}}

\def\go{G_Z^0}
\def\gp{{G^+_W}}

\def\mhpm{m_{\hpm}}

\def\mt{m_t}
\def\mb{m_b}

\def\ls#1{\ifmath{_{\lower1.5pt\hbox{$\scriptstyle #1$}}}}

\begin{document}

\title{Bounds on charged higgs boson in the 2HDM type III from Tevatron}
\author{R. Martinez} 
\author{J-Alexis Rodriguez } 
\author{M. Rozo}
\affiliation{Departamento de Fisica, Universidad Nacional de Colombia\\
Bogota, Colombia}



\begin{abstract}
We consider the Two Higgs Doublet Model (2HDM) of type III which leads to
Flavour Changing Neutral Currents (FCNC) at tree level. In the framework of this model we can  use an appropriate form of the Yukawa Lagrangian that makes the type II model limit of the general type III couplings apparent. This way is useful in order to compare with the experimental data which is model dependent. The analytical expressions of the partial  width $\Gamma \left( t \rightarrow H^+ b \right) $ are derived and we compare with the data available at this energy range. We examine the limits on the new parameters $\lambda_{ij}$ from the validness of perturbation theory. 
\end{abstract}

\maketitle

The Standard Model (SM) of particle physics based on the gauge group $SU(3)_c \times SU(2)_L \times U(1)_Y$ accommodates the symmetry breaking by including a fundamental weak doublet of scalar Higgs bosons $\phi$ with a scalar potential  $V(\phi)= \lambda (\phi^\dagger \phi - \frac 12 v^2)^2$. However, the SM does not explain the dynamics responsible for the generation of masses. Furthermore, the scalar sector suffers from two serious problems, known as: the gauge hierarchy problem and the triviality problem \cite{hunter}. The scalars involved in electroweak symmetry breaking should therefore be a party to new physics at some finite energy scale. Thus the SM would be merely a low-energy effective field theory, and the dynamics responsible for generating mass might lie in physics beyond the SM. There is the option of a model like the SM but including a richer scalar sector, which includes one more Higgs doublet, it is  called generically the Two Higgs Doublet Model (2HDM).

 There are several kinds of such 2HDM models. In the model called type I, one Higgs Doublet provides masses to the up and down quarks, simultaneously. In the model type
II, one Higgs doublet gives masses to the up quarks and the other one to the
down quarks. These two models have a discrete symmetry  to avoid FCNC at tree level \cite{gw}. However, the discrete symmetry
is not necessary in whose case both doublets generate the masses of the
quarks of up-type and down-type, simultaneously. In the literature, the
latter model is known as the model type III \cite{III}. It has been used to look for physics beyond the SM and specifically for FCNC at tree level \cite{ARS,Sher, we}. In general, both doublets could acquire a vacuum
expectation value (VEV), but  one of them can be absorbed redefining the Higgs
fields properly. Nevertheless, we have showed that from the case in which both doublets get the VEV is possible to study the models type I and II in an specific limit \cite{we}. Therefore we
consider the model type III in  two basis. In the first base, the
two Higgs doublets acquire VEV (case (a) in ref.\cite{we}). In the second one, only one Higgs
doublet acquire VEV (case (b) in ref\cite{we})\cite{ARS}. In the latter case the free parameter $\tan
\beta\equiv v_2/v_1 $ is removed from the theory making its phenomenological analysis simpler. But in the former one is possible to get bounds for the model type III using the experimental bounds which have been gotten in the framework of the model type II.

In these kind of models (2HDM) additional degrees of freedom appear, providing a total of five observable Higgs fields: two neutral CP-even scalars $h^0$ and $H^0$, a neutral CP-odd scalar $A^0$, and two charged scalars $H^\pm$. Direct searches have carried out by LEP experiments, and report a combined lower limit on $M_{H^\pm}$ of 78.6 GeV \cite{lep}. The CDF collaboration has also reported a direct search for charged Higgs boson, setting an upper limit on $B(t \to H^+b)$ around  0.6 at 95 \% C.L. for masses in the range 60-160 GeV \cite{cdf}. On the other hand, indirect and direct searches have been carried out by D0 looking for a decrease  in the $t \bar t \to W^+ W^- b \bar b$ signal expected from the SM and the direct search for the decay mode $H^\pm \to  \tau^\pm \nu$. They exclude most regions of the plane $M_{H^\pm}-\tan \beta$ where the $B(t \to b H^+)>0.36$ \cite{d0}. We should note that all the bounds have been gotten in the framework of the 2HDM type II. And, in the framework of the 2HDM type II and MSSM a full one loop calculation of $\Gamma (t \to b H^+)$ including all sources for large Yukawa couplings were presented in references \cite{qcd,otros}. In what follows we concentrate on the charged sector, with the relevant parameters being its mass $M_{H^\pm}$ and the ratio of the VEV's of the doublets, $\tan \beta$ and the coupling intensities  $\lambda_{tt}$ and $\lambda_{bb}$.  

In the present work, we study the process $t \rightarrow b H^+$ in the 2HDM type III. If $m_{H^\pm}<m_t-m_b$ then the charged Higgs boson $H^\pm$ can be produced in the decay of the top quark via $t \to bH^+$. This decay can be competitive with the dominant SM decay mode, $t \to b W^+$. The Higgs boson production in top decays has been studied in the framework of the 2HDM type II and under considerations that also apply to the MSSM \cite{otros}. We are going to work in the Higgs mass range 60-160 GeV, assuming that $B(t \to b W^+)+B(t \to b H^+)=1 $ and the masses of the neutral scalars are assumed to be large enough to be suppressed in $H^\pm$ decays. In this way the only available decays of $H^\pm$ are fermionic.

The 2HDM type III is an extension of the SM plus a new Higgs doublet and three new Yukawa couplings in the quark and leptonic sectors. The mass terms for the up-type or down-type sector depends on two matrices or two Yukawa couplings. The rotation of the quarks and lepton gauge eigenstates allow us to diagonalize one of the matrices but not both simultaneously, so one of the Yukawa couplings remains non-diagonal, generating the FCNC at tree level.

The Higgs couplings to fermions are model dependent. The most general structure for the Higgs-fermion Yukawa couplings, 2HDM type-III  \cite{III}, is as follow:
\begin{eqnarray}
-\pounds _{Y} &=&\eta _{ij}^{U,0}\overline{Q}_{iL}^{0}\widetilde{\Phi }%
_{1}U_{jR}^{0}+\eta _{ij}^{D,0}\overline{Q}_{iL}^{0}\Phi _{1}D_{jR}^{0}+\eta
_{ij}^{E,0}\overline{l}_{iL}^{0}\Phi _{1}E_{jR}^{0} \nonumber \label{Yukawa} \\
&+&\xi _{ij}^{U,0}\overline{Q}_{iL}^{0}\widetilde{\Phi }_{2}U_{jR}^{0}+\xi
_{ij}^{D,0}\overline{Q}_{iL}^{0}\Phi _{2}D_{jR}^{0}+\xi _{ij}^{E,0}\overline{%
l}_{iL}^{0}\Phi _{2}E_{jR}^{0}\nonumber \\
&+& h.c. 
\end{eqnarray}
where $\Phi _{1,2}\;$are the Higgs doublets, $\widetilde{\Phi}_i\equiv
i\sigma_2 \Phi^*_i$,
$Q_L^0 $ is the weak isospin quark doublet, 
and $U^0_R$, $D^0_R$ are weak isospin quark singlets, whereas $\;\eta _{ij}^{0}\;$and $\xi
_{ij}^{0}\;$ are non-diagonal $3\times 3\;$non-dimensional matrices and $i$, 
$j$ are family indices. The superscript $0$ indicates that the fields are
not mass eigenstates yet. In the so-called model type I, the discrete
symmetry forbids the terms proportional to $\eta _{ij}^{0},\;$meanwhile in
the model type II the same symmetry forbids terms proportional to $\xi
_{ij}^{D,0},\;\eta _{ij}^{U,0},\xi _{ij}^{E,0}.$
We next shift the scalar fields according to their VEV's, as 
\begin{equation}
\left\langle \Phi _{1}\right\rangle _{0}=\left( 
\begin{array}{c}
0 \\ 
v_{1}/\sqrt{2}
\end{array}
\right) \;\;,\;\;\left\langle \Phi _{2}\right\rangle _{0}=\left( 
\begin{array}{c}
0 \\ 
v_{2}/\sqrt{2}
\end{array}
\right)  \nonumber
\end{equation}
and we take the complex phase of $v_{2}\;$equal to zero since we are not
interested in CP violation. Then re-express the scalars in terms of the physical Higgs
states and would-be Goldstone bosons,
\begin{eqnarray}
\left( 
\begin{array}{c}
G_{W}^{\pm } \\ 
H^{\pm }
\end{array}
\right) &=&\left( 
\begin{array}{cc}
\cos \beta & \sin \beta \\ 
-\sin \beta & \cos \beta
\end{array}
\right) \left( 
\begin{array}{c}
\phi _{1}^{\pm } \\ 
\phi _{2}^{\pm }
\end{array}
\right) ,  \nonumber \\
\left( 
\begin{array}{c}
G_{Z}^{0} \\ 
A^{0}
\end{array}
\right) &=&\left( 
\begin{array}{cc}
\cos \beta & \sin \beta \\ 
-\sin \beta & \cos \beta
\end{array}
\right) \left( 
\begin{array}{c}
\sqrt{2}Im\phi _{1}^{0} \\ 
\sqrt{2}Im\phi _{2}^{0}
\end{array}
\right) ,  \nonumber \\
\left( 
\begin{array}{c}
H^{0} \\ 
h^{0}
\end{array}
\right) &=&\left( 
\begin{array}{cc}
\cos \alpha & \sin \alpha \\ 
-\sin \alpha & \cos \alpha
\end{array}
\right) \left( 
\begin{array}{c}
\sqrt{2}Re\phi _{1}^{0}-v_{1} \\ 
\sqrt{2}Re\phi _{2}^{0}-v_{2}
\end{array}
\right)  \label{Autoestados masa Higgs}
\end{eqnarray}
where $\tan \beta \equiv t_\beta=v_{2}/v_{1}\;$and $\alpha \;$is the mixing angle of the $h^0$ , $H^0$ 
CP-even neutral Higgs sector. $G^{0(\pm)}_{Z(W)}\;$are the would-be Goldstone bosons
for $Z^0\left( W^\pm \right)$, respectively. And $A^{0}\;$is the CP-odd neutral
Higgs. $H^{\pm }\;$are the charged physical Higgses.
  
In addition, we diagonalize the quark
mass matrices and define the quark mass eigenstates.
The resulting Higgs-fermion Lagrangian can be written in
several ways \cite{we}.  We choose to display the form that
makes the type-II model limit 
of the general type-III couplings apparent.
In the model type-II  (where $\eta^{U,0}_{ij}=\xi^{D,0}_{ij}=0$)  tree-level Higgs mediated flavor-changing neutral currents are automatically absent, whereas
these are generally present for type-III couplings.
The fermion mass eigenstates are related to the interaction eigenstates
by biunitary transformations:
\beqa
&& U_L =V_L^U  U^0_L\,,\qquad U_R =V_R^U  U^0_R\,,\nonumber \\
&& D_L =V_L^D  D^0_L\,,\qquad D_R =V_R^D  D^0_R\,,
\eeqa
and the Cabibbo-Kobayashi-Maskawa matrix is defined as $K\equiv V_L^U
V_L^{D\,\dagger}$.  It is also convenient to define ``rotated''
coupling matrices:
\beqa
\eta^U (\xi^U) & \equiv & V_L^U \eta^{U,0}(\xi^{U,0}) V_R^{U\,\dagger}\,,\nonumber \\
\eta^D (\xi^D) & \equiv & V_L^D \eta^{D,0}(\xi^{D,0}) V_R^{D\,\dagger}\,.
\eeqa
The diagonal quark mass matrices are obtained by replacing the scalar
fields with their VEV's:
\begin{equation}
\mdd={1\over \sqrt 2}(v_1\eta^D+v_2\xi^D)\,,\quad
\mud={1\over \sqrt 2}(v_1\eta^U+v_2\xi^U)\,.
\end{equation}
After eliminating $\eta^{D}\,, \, \xi^U$, the 
resulting Yukawa couplings are \cite{we}:
\begin{widetext}
\beqa
\call_Y&=&
{1\over v}\,\anti D\mdd D\left({\sa\over \cb}\hl-{\ca\over \cb}\hh\right)
+{i\over v}\,\anti D\mdd\gamma_5 D(\tb\ha-\go)\nonumber \\
&&-{1\over \rtwo\cb}\anti D (\xi^D P_R+ {\xi^D}^\dagger P_L)D
(\cbma\hl-\sbma\hh)
-{i\over \rtwo\cb}\anti D (\xi^D P_R- {\xi^D}^\dagger P_L)D\,\ha
\nonumber\\
&&-{1\over v}\,\anti U\mud U\left({\ca\over \sb}\hl+{\sa\over \sb}\hh\right)
+{i\over v}\,\anti U\mud\gamma_5 U(\tbi\ha+\go)\nonumber\\
&&+{1\over \rtwo\sb}\anti U (\eta^U P_R+ {\eta^U}^\dagger P_L)U 
(\cbma\hl-\sbma\hh)
-{i\over \rtwo\sb}\anti U (\eta^U P_R- {\eta^U}^\dagger P_L)U\,\ha 
\nonumber\\
&&+{\rtwo\over v}\biggl[\anti U K\mdd P_R D(\tb\hp-\gp)
+\anti U \mud K P_LD(\tbi\hp+\gp)+{\rm h.c.}\biggr]\nonumber\\
&&-\left[{1\over \sb}\anti U {\eta^U}^\dagger K P_LD\,\hp
+{1\over \cb}\anti U K\xi^D  P_RD\,\hp +{\rm h.c.}\right]\,.
\label{yukform}
\eeqa
\end{widetext}
where we have used the notation $s(c)_{\alpha}=\sin (\cos)\alpha$ and $\sin(\beta-\alpha)=s_{\beta-\alpha}$ and so on.

In the 2HDM type III after using the parameterization proposed by Cheng and Sher \cite{Sher} for the couplings $\xi(\eta)_{ii}=\lambda_{ii} g m_i/(2m_W)$, we get the following expression for the decay width $t \to b H^+$,
\begin{eqnarray}
\Gamma(t \to b H^+) =\frac{G_F  K_{tb}^2}{4 \pi  \sqrt{2}} &&\left[ a^2 (m_t^2+m_b^2-m_{H^\pm}^2)+4 ab m_t m_b \right.\nonumber \\
&+&\left. b^2 (m_t^2+m_b^2 -m_{H^\pm}^2)\right] \vert \vec p_H \vert
\end{eqnarray}
where
\begin{eqnarray}
\vert \vec p_H \vert&=& \left[(m_t^2-(m_b+m_{H^\pm})^2)(m_t^2-(m_{H^\pm}-m_b)^2)\right]^{1/2}/(2m_t) \,,\\
a &=& \cot \beta -\frac{\lambda_{tt}}{\sqrt{2}}\csc \beta \, ,\nonumber \\
b &=&\frac{m_b}{m_t}\left( \tan \beta - \frac{\lambda_{bb}}{\sqrt{2}\cos \beta}\right)\, .
\end{eqnarray}
Further we have taken the products $(\eta K)_{33}\sim \eta_{tt} K_{tb}$ and $(K \xi)_{33}\sim \xi_{bb} K_{tb}$, neglecting the off-diagonal terms because they are suppressed by the CKM entries. From the  expression  (8) is possible to get the decay width in the framework of the 2HDM-II just replacing $\lambda_{ii}=0$.

   In order to proceed with the numerical evaluations, we wonder about the perturbation regime. First of all, we are calculating a decay width at tree level, therefore we should take into account the possible values of $\lambda$ which should be consistent with perturbation theory. Looking at the coupling $\bar t  b H^+$ from (7), we get 
\beq
\frac{\mb^2}{\mt^2} \left \vert \rtwo \frac{t_\beta}{\sqrt{1+t_\beta^2}}-\lambda_{bb} \right \vert^2 + t_\beta^{-2} \left \vert  \frac{\rtwo}{\sqrt{1+t_\beta^2}}-\lambda_{tt} \right \vert^{2} < \frac{8}{1+t_\beta^2}\, .
\eeq
The allowed region in the plane $\lambda_{tt}-\lambda_{bb}$ depends on $t_\beta$, but for a wide range of $t_\beta$, $\lambda_{bb}$ is inside the interval (-100 , 100) while $\lambda_{tt}$ is in (-2.8 , 2.8). 

 On the other hand,  we can consider the 2HDM-III in a basis where only one Higgs doublet acquire VEV and then it does not have the parameter $\tan \beta$. It is the usual 2HDM type III \cite{Sher}, where the Lagrangian of the charged sector is given by
\beq
-L^{III}_{H^\pm ud}=H^+ \anti U [ K \xi^D P_R-\xi^U K P_L] D+h.c.
\eeq
With the above Lagrangian the model is simpler due to the absence of the $\tan \beta$ parameter, and it can be obtained from a Lagrangian written  in  a basis where both doublets acquire VEV different from zero \cite{we}. 
For the decay width (8), it is reduced because now we have $a=\frac{\lambda_{tt}}{\sqrt{2}}$ and $b=\frac{\lambda_{bb}m_b}{\sqrt{2}m_t}$. And from the perturbation theory consideration we get
\beq
\frac{\mb^2}{\mt^2} \vert \lambda_{bb} \vert^2 + \vert \lambda_{tt} \vert^2 <8 ,\label{uno}
\eeq 
which is an ellipse with $\vert \lambda_{bb}\vert \leq 100$ and $\vert \lambda_{tt} \vert \leq \sqrt{8}$; this constraint might be satisfied. In this context, taking into account the bound from D0 experiment, $B(t \to H^+ b)\leq 0.36$  \cite{d0} we present in figure 1 the plane $\mhpm$ vs $\lambda_{bb}$. We are using the constraint from perturbation in order to get the upper limit allowed in this plane. The allowed region is above the curve shown. For small values of $\lambda_{bb}$ the charged Higgs boson mass has a lower bound of  $\sim 140 GeV$. 

But notice that the bounds coming from the experiment (LEP and Tevatron) are really gotten in a model dependent way, specifically they have been gotten in the framework of the 2HDM-II. Then we should use the parameterization given by the Lagrangian (\ref{yukform}) which is the 2HDM-II plus new changing flavor interactions. This reason makes useful the Lagrangian (\ref{yukform}) in order to do comparisons of 2HDM-III with the experimental values obtained using the 2HDM-II. In figures 2 and 3, we show the fraction $B(t \to b H^+)$ vs $\tan \beta$ for  different values of $\lambda_{bb}$ and $\lambda_{tt}$. The  $B(t \to b H^+)$ branching fraction for different values of $M_{H^\pm}$ is significant for very small or very large values of $\tan \beta$, while it is suppressed for intermediate values of $\tan \beta$. This fact is because the fraction $B(t \to b H^+)$ is proportional to $(m_b^2+m_t^2-m_{H^\pm})(m_t^2 \cot^2 \beta+m_b^2 \tan^2 \beta)+4 m_t^2 m_b^2$ which has a minimum around $\tan \beta =\sqrt{m_t/m_b}$ and is symmetric in $\log(\tan \beta)$ about this point. We note that for  a charged Higgs 2HDM-II like, lighter than the top quark, it would be detected if $\tan \beta$ is substantially different from $\sqrt{m_t/m_b}$, it is because  the branching fraction is very suppressed around this point, by around $10^{-4}$.  In what follows we are going to show that model type III can modify this scenario.  For $\lambda_{ii}=0$ we obtain the prediction of the 2HDM type II and it is symmetric around $\tan \beta=6$, but it is not the case for $\lambda_{ii} \neq 0$ where the minimum is shifted.  In this analysis  we are considering two different set of values for the $\lambda_{ii}$ parameters. In figure 2, $\lambda_{tt}$ is the order of $\lambda_{bb}$ and in figure 3 $\lambda_{tt}$ is one order of magnitude smaller than $\lambda_{bb}$, and  in both cases the charged Higss boson mass is fixed to 140 GeV.  In figure 2 and 3, it is drawn the most stringent bound coming from D0 collaboration on $B(t \to b H^+)$ in the range of $0.3 \leq \tan \beta \leq 150$ which should be less than 0.36 (horizontal line).

Finally, in figure 4 we plot $M_{H^\pm}$ vs $\tan \beta$ for different $\lambda_{ii}$ using the bound from Tevatron $B(t \to b H^+)\leq 0.36$. We show the excluded regions at 95 \% C.L. by Tevatron from Run I and the limits that will be reached on this plane in Run II using 2 and 10 fb$^{-1}$ for the integrated luminosity at $\sqrt{s}=2$ TeV. We should clarify that the exclusion regions taken from D0 at Tevatron are a combination of two searches. An indirect search, looking for a decrease in $t \bar t \to W^+ W^- b \bar b$ signal expected from the SM, this search excludes simultaneously both large and small $\tan \beta$.  And a direct search, that look for  the $H^\pm \to \tau^\pm \nu$ in the region $0.3 < \tan \beta < 150$. This is because the fraction rate of the leptonic decay of $H^\pm$ is around 0.95 for large $\tan \beta$, and the other option $H^+ \to c \bar s$ is important for $\tan \beta< 0.4$ and low Higgs boson mass. In our case we could enhanced the influence of $H^+ \to c \bar s$ channel  due to the appearance of new couplings, it does not matter the value of $\tan \beta$, but that possibility corresponds to a very unusual set of values of the new couplings $\lambda_{cc}$, $\lambda_{ss}$ and $\xi^E$, instead of that we conserve the hierarchy of the decay channels according to the value of $\tan \beta$ in order to consider a more general and conservative scenario where the experimental limits can be used. The solid line inside the future explored region corresponds to $\lambda_{ii}=0$ which is the region for the 2HDM-II with a fraction rate of 0.36. A charged Higgs 2HDM-III like could have  different scenarios, as we can see from figure 4 for $\lambda_{ii}$ different from zero. Again we are considering two cases: $\lambda_{tt}$ of the order of $\lambda_{bb}$ and $\lambda_{tt}$ one order of magnitude smaller than $\lambda_{bb}$. The excluded region is below the curves, and we can see that there are values that cover almost all the plane presented.

To summarize, in the present work we have examined a 2HDM type III which produces FCNC at tree
level, in general these new interactions are governed by parameters $\lambda_{ij}$.  The experimental analysis has been carried out using the 2HDM type II as a framework, so they are model dependent. We have already presented a form of the Yukawa Lagrangian of the 2HDM-III (\ref{yukform}) which can be reduced to the 2HDM-II as a limit \cite{we}, in this way it can be used to compare the model type III with the experimental analysis based on 2HDM type II.
 We have shown that 2HDM type III can modify the situation for the branching fraction $B(t \to b H^+)$ in the case of a charged Higgs boson in the allowed region from kinematic  considerations. For $\lambda_{ii}=0$, we obtain the prediction of the 2HDM type II and it is symmetric around $\tan \beta=\sqrt{m_t/m_b} \sim 6$, but it is not the case for $\lambda_{ii} \neq 0$. Finally, we have presented the parameter plane $\tan \beta-m_{H^+}$ showing the experimental limits for Run I and II and, in the same plot we show the solutions obtained in the cases of 2HDM-II ($\lambda_{ii}=0$) and 2HDM-III ($\lambda_{tt} \neq 0$).

We acknowledge to R.A. Diaz  for the careful
reading of the manuscript, and J. P. Idarraga for his collaboration with numerical calculations. This work was supported by COLCIENCIAS.

\newpage

\begin{figure}[htbp]
\begin{center}
\includegraphics[angle=0,width=10cm]{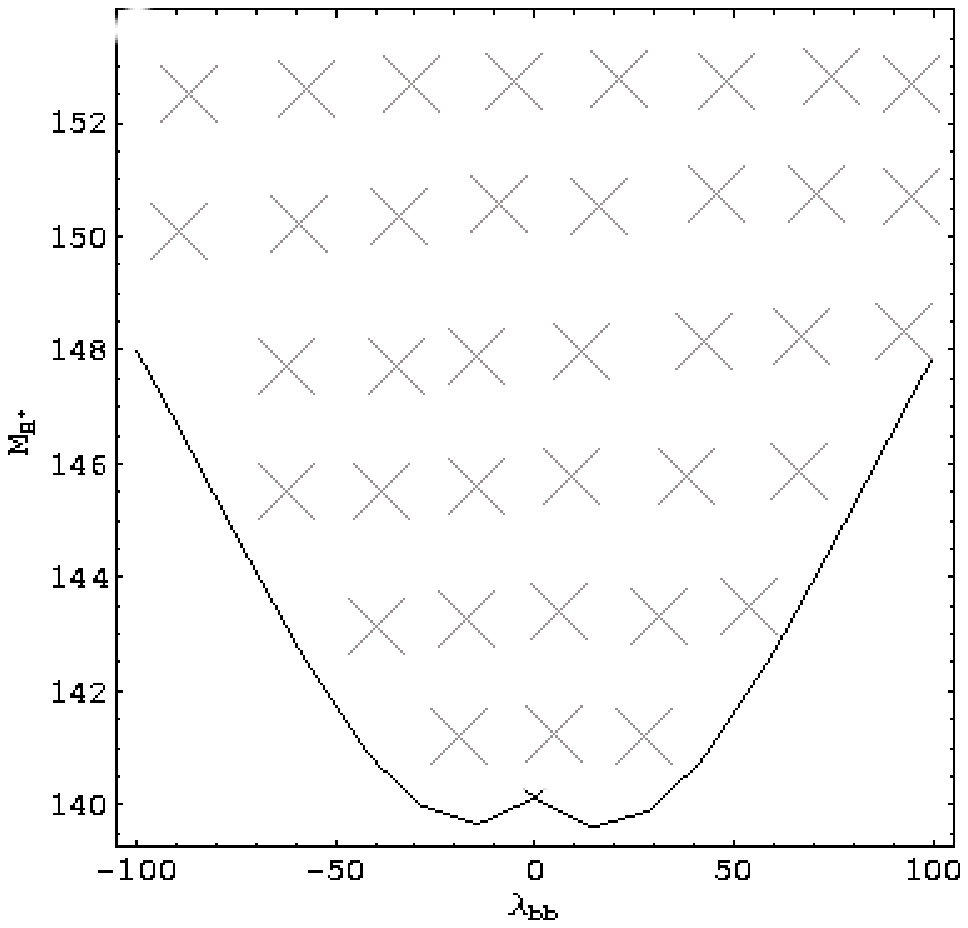}
\end{center}
\caption{Plot for the charged Higgs boson mass $M_{H_\pm}$ versus the parameter $\lambda_{bb}$. In the 2HDM-III when $\tan \beta$ is not present. The allowed region is above the curve.}
\end{figure}

\begin{figure}[htbp]
\begin{center}
\includegraphics[angle=0,width=15cm]{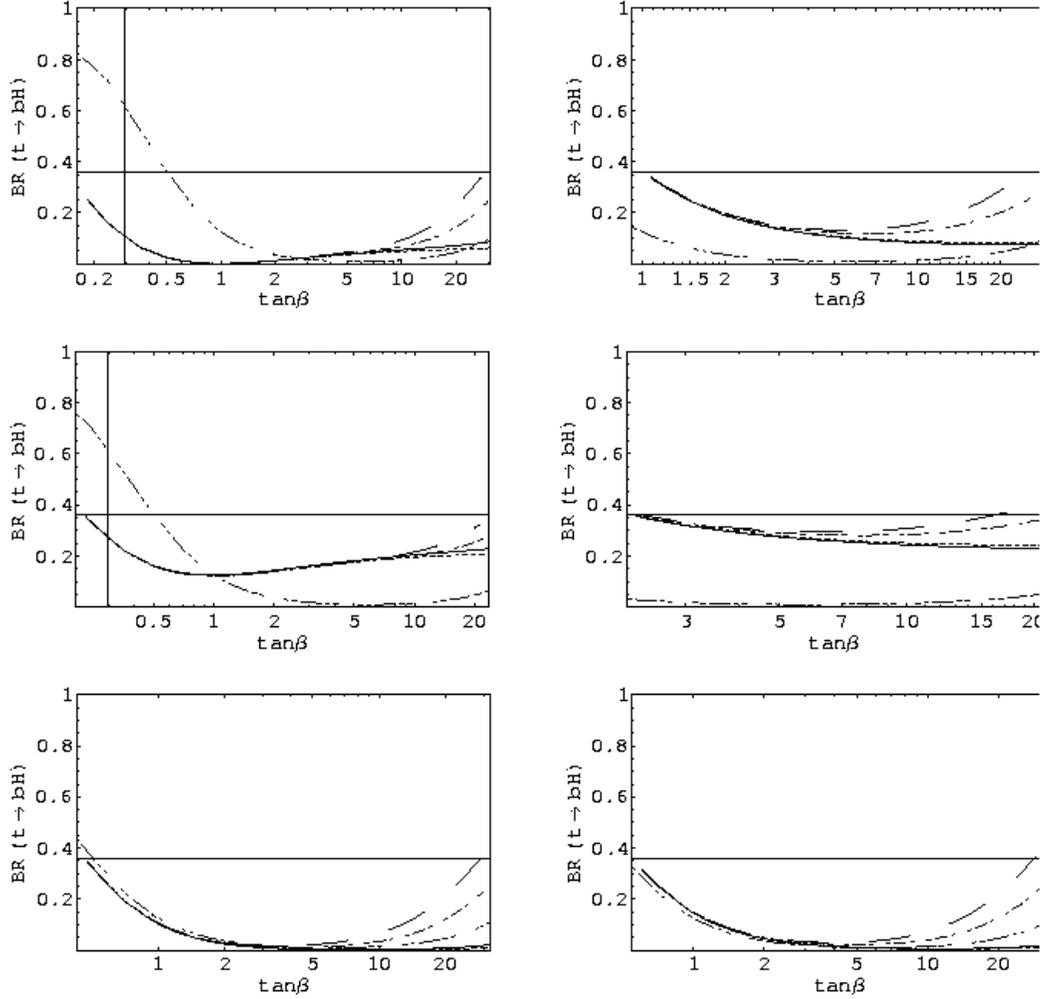}
\end{center}
\caption{ Plot for $B(t \to b H^+)$ versus the $\tan \beta$ parameter for  $M_{H_\pm}=140$ GeV and
different values of the parameter $\lambda_{ii}$. Each plane corresponds to a fixed $\lambda_{tt}$, downward $=\pm 1,\pm 2,\pm 0.1$ right side positive values, left side negative values. And  $\lambda_{bb}=-2$ dashed-line, $=-1$  dot-dashed line  $=1$ short-dashed line, and $=2$ solid line. We also show the 2HDM-II case, dot-dot-dashed line, showing the minimum around $\tan \beta \sim 6$. The excluded region by Tevatron (horizontal line) corresponds to the right-up corner and the region for $\tan \beta<0.3$ (vertical line) is an unexplored region by experiments.}
\end{figure}

\begin{figure}[htbp]
\begin{center}
\includegraphics[angle=0,width=15cm]{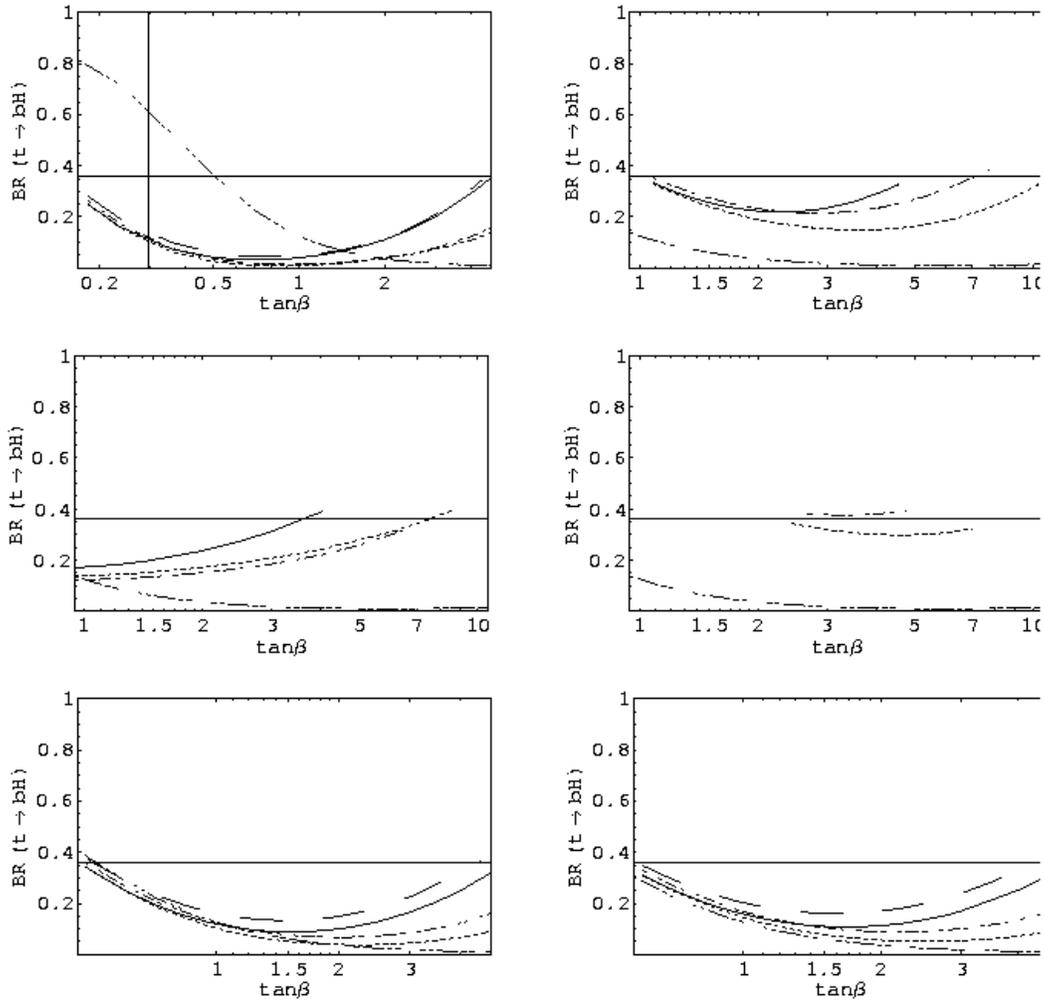}
\end{center}
\caption{ Like figure 2 but $\lambda_{bb}=-20$ dashed-line, $=-10$  dot-dashed line  $=10$ short-dashed line, and $=20$ solid line.}
\end{figure}

\begin{figure}[htbp]
\begin{center}
\includegraphics[angle=0,width=10cm]{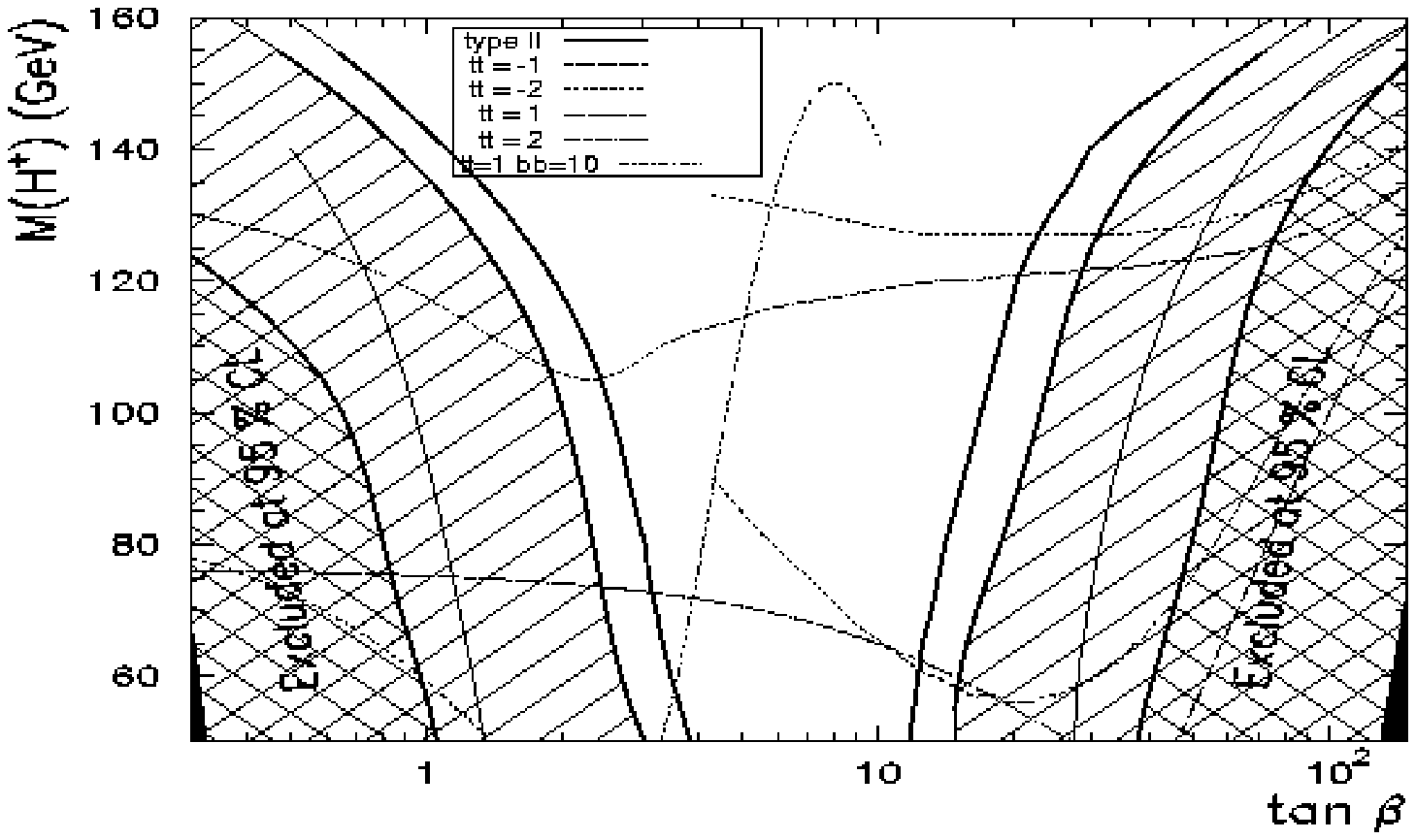}
\end{center}
\caption{ Plot for $M_{H^\pm}$ versus the $\tan \beta$ parameter when $B(t \to b H^+) \leq 0.36$ for different set of values of $\lambda_{ii}$. Inside the figure are the labels for $\lambda_{tt}$ and $\lambda_{bb}=1$ except for the last one which is $\lambda_{bb}=10$. We overlap the expected limits from D0 for $m_t=175$ GeV and several values of the integrated luminosity and $\sqrt{s}$: (0.1 fb$^{-1}$, 1.8TeV),(2 fb$^{-1}$, 2 TeV),(10 fb$^{-1}$, 2 TeV), assuming  $\sigma(t \bar t)=7$ pb. The limits were taken from reference \cite{otros}.}
\end{figure}


\newpage
\begin{thebibliography}{000}

\bibitem{hunter}  For a review see J. Gunion, H. Haber, G. Kane and S. Dawson, 
\textit{The Higgs Hunter's Guide}, (Addison-Wesley, New York, 1990)

\bibitem{gw}  S. Glashow and S. Weinberg, Phys. Rev. D \textbf{15}, 1958
(1977).

\bibitem{III}  W.S. Hou, Phys. Lett B \textbf{296}, 179 (1992); D. Cahng, W.
S. Hou and W. Y. Keung, Phys. Rev. D \textbf{48}, 217 (1993); S. Nie and M. Sher, Phys. Rev. D \textbf{58}, 097701 (1998); M. Sher and Y. Yuan, Phys. Rev. D \textbf{44}, 1461 (1991); D. Atwood, L. Reina and A. Soni, Phys. Rev. D \textbf{55}, 3156 (1997); Phys. Rev. Lett. \textbf{75}, 3800 (1995).

\bibitem{ARS}  D. Atwood, L. Reina and A. Soni, Phys. Rev. D \textbf{53},
1199 (1996); Phys. Rev. D \textbf{54}, 3296 (1996); Phys. Rev. Lett. 
\textbf{75}, 3800 (1993); D. Atwood, L. Reina and A. Soni, Phys. Rev. D \textbf{55}, 3156 (1997); G. Cvetic, S. S. Hwang and C. S. Kim, Phys. Rev. D \textbf{58}, 116003 (1998).
\bibitem{Sher}  Marc Sher and Yao Yuan, Phys. Rev. D \textbf{44}, 1461
(1991); T.P. Cheng and M. Sher, Phys. Rev. D \textbf{35}, 3490
(1987)
\bibitem{we} Rodolfo A. Diaz, R. Martinez and J.-Alexis Rodriguez, Phys. Rev. D \textbf{64}, 033004 (2001); Phys. Rev. D \textbf{63},  095500 (2001).
\bibitem{lep} LEP Collaborations, arXiv:hep-ex/0107031.
\bibitem{cdf} CDF Collaboration, F. Abe,\,\textit{et. al}, Phys. Rev. Lett. \textbf{79}, 357 (1997); CDF Collaboration, T. Affolder, \textit{et. al}, Phys. Rev. D \textbf{62}, 012004 (2000).
\bibitem{d0} D0 Collaboration, B. Abbot, \textit{et. al}, Phys. Rev. Lett. \textbf{82}, 4975 (1999); V. Abazov, \textit{et. al}, Phys. Rev. Lett. 88, 151803 (2002).
\bibitem{qcd} A. Mendez and A. Pomarol, Phys. Lett. B \textbf{360}, 47 (1995); C. Li and R. J. Oakes, Phys. Rev. D \textbf{43}, 855 (1991); A. Djouadi and P. Gambino, Phys. Rev. D \textbf{51}, 218 (1995). 
\bibitem{otros}M. Carena, D. Garcia, U. Nierste, C. Wagner, arXiv:hep-ph/9912516; M. Carena, J. Conway, H. Haber and J. Hobbs, arXiv:hep-ph/0010338. 

\end{thebibliography}
\end{document}